\documentstyle[amstex,preprint,aps]{revtex}

\begin{document}
\title{Exact result on the supercurrent through a
superconductor/quantum-dot/superconductor junction}
\author{Wei Li , Yu Zhu, and Tsung-han Lin}
\address{{\it State Key Laboratory for Mesoscopic Physics and }\\
{\it Department of Physics, Peking University,}{\small \ }{\it Beijing}\\
100871, China}
\maketitle

\begin{abstract}
We present an analytical result for the supercurrent across a
superconductor/quantum-dot/superconductor junction. By converting the
current integration into a special contour integral, we can express the
current as a sum of the residues of poles. These poles are real and give a
natural definition of the Andreev bound states. We also use the exact result
to explain some features of the supercurrent transport behavior.
\end{abstract}


PACS numbers: 74.50.+r, 73.40.Gk, 73.20.-b, 73.63.Kv.

\baselineskip 20pt 

\smallskip

{\it Introduction.} The mesoscopic superconductor/quantum-dot/superconductor
(S-QD-S) junction is a typical structure to study the phase coherent current
transport in mesoscopic hybrid systems \cite{Yeyati1997}. When using Keldysh
Green function technique \cite{Keldysh1964} to evaluate the current across
the S-QD-S junction, the usual method demands a numerical integration and
faces the problem of explaining some features of the current transport
behavior \cite{Zhu2001}. In this paper, we employ a new method to
analytically compute the current across the S-QD-S junction in the absence
of voltage bias, and use the analytical results to explain some features of
current transport properties.

{\it Model and Hamiltonian.} In the system under consideration, a quantum
dot defined in a 2-dimensional electron gas (2DEG) is coupled to two BCS
superconducting leads. We model this system by the Hamiltonian 
\begin{equation}
\hat{H}=\hat{H}_L+\hat{H}_d+\hat{H}_R+\hat{H}_T\text{ ,}
\end{equation}
where 
\begin{equation}
\hat{H}_L=\sum_{k,\sigma }\epsilon _k^La_{k\sigma }^{\dagger }a_{k\sigma
}+\sum_k[\Delta _La_{k\uparrow }^{\dagger }a_{-k\downarrow }^{\dagger
}+\Delta _L^{*}a_{-k\downarrow }a_{k\uparrow }]
\end{equation}
\begin{equation}
\hat{H}_d=\epsilon _0(d_{\uparrow }^{\dagger }d_{\uparrow }+d_{\downarrow
}^{\dagger }d_{\downarrow })\text{ ,}
\end{equation}
\begin{equation}
\hat{H}_R=\sum_{p,\sigma }\epsilon _p^Rb_{p\sigma }^{\dagger }b_{p\sigma
}+\sum_p[\Delta _Rb_{p\uparrow }^{\dagger }b_{-p\downarrow }^{\dagger
}+\Delta _R^{*}b_{-p\downarrow }b_{p\uparrow }]\text{ ,}
\end{equation}
are the isolated (unperturbed) Hamiltonians of the left superconductor, the
quantum dot, and the right superconductor, respectively.

We consider only a quantum dot with a negligible intra-dot Coulomb
interaction. The reason is that we want to focus on the superconducting
proximity effect on the QD, which is due to Andreev reflections; whereas a
large charging effect caused by intra-dot Coulomb repulsion suppresses the
transport mediated by Andreev process \cite{Ralph1997,Kang1998}. Study on
the competition of a medium intra-dot Coulomb interaction with Andreev
reflections in S-QD-S junction will be carried on in the coming work.
Furthermore, we consider only one single energy level inside the QD. The
generalization to QDs with several energy levels is straightforward.

The tunneling term 
\begin{equation}
\hat{H}_T=\sum_{k,\sigma }[L_ka_{k\sigma }^{\dagger }d_\sigma
+c.c.]+\sum_{p,\sigma }[R_pb_{k\sigma }^{\dagger }d_\sigma +c.c.]\text{ ,}
\end{equation}
describes the electron-transfer between the QD and the leads.

\bigskip

{\it Current Formula.} The current flowing across the junction is \cite
{Sun2000} 
\begin{equation}
I=-\frac{2e\Gamma }\hbar \text{Tr}\{%
\mathop{\rm Re}%
\int_{-\infty }^{+\infty }\frac{d\omega }{2\pi }\left[ {\bf G}_d^r\left(
\omega \right) {\bf g}_L^{<\text{ }}\left( \omega \right) +{\bf G}%
_d^{<}\left( \omega \right) {\bf g}_L^a\left( \omega \right) \right] \}\text{
,}
\end{equation}
where ${\bf g}_L$, ${\bf G}$ are the Keldysh Green functions of the left
superconductor and the QD in Nambu space \cite{Keldysh1964}. The retarded
Green function of the superconductor is 
\begin{equation}
{\bf g}_{L/R}^r\left( \omega \right) =(-i\pi )\rho _{L/R}(\omega
+i0^{+})\left( 
\begin{array}{ll}
1 & \frac{\Delta _{L/R}}\omega  \\ 
\frac{\Delta _{L/R}^{*}}\omega  & 1
\end{array}
\right) \text{ ,}
\end{equation}
where $\rho _{L/R}(\omega )=\frac{\left| \omega \right| }{\sqrt{\omega
^2-\Delta _{L/R}^2}}$ is the density of states (DOS) of a BCS
superconductor. After solving the matrix Dyson equation, we obtain the
retarded Green function of the QD:

\begin{equation}
{\bf G}_{d}^{r}\left( \omega \right) =\frac{1}{D(\omega )}\left( 
\begin{array}{cc}
g^{r}\left( \omega \right) _{22}^{-1}-\Sigma ^{r}\left( \omega \right) _{22}
& \Sigma ^{r}\left( \omega \right) _{12} \\ 
\Sigma ^{r}\left( \omega \right) _{21} & g^{r}\left( \omega \right)
_{11}^{-1}-\Sigma ^{r}\left( \omega \right) _{11}
\end{array}
\right) \text{ ,}
\end{equation}
where 
\begin{equation}
{\bf g}^{r}\left( \omega \right) =\left( 
\begin{array}{cc}
\frac{1}{\omega -\epsilon _{0}+i0^{+}} & 0 \\ 
0 & \frac{1}{\omega +\epsilon _{0}+i0^{+}}
\end{array}
\right) \text{ }
\end{equation}
is the unperturbed retarded Green function of the QD. The QD's self-energy

\begin{equation}
{\bf \Sigma (\omega )=}\frac \Gamma {2\pi }{\bf g}_L\left( \omega \right) +%
\frac \Gamma {2\pi }{\bf g}_R\left( \omega \right) \text{ ,}
\end{equation}
describes the proximity effect of superconductors on the QD; here we have
used the wide band-width approximation \cite{Wingreen1989}. $D(\omega )$ in
Eq. [8] is the determinant of the matrix $({\bf g}^{r-1}-{\bf \Sigma }^r{\bf %
)}$: 
\begin{equation}
D(\omega )=[g^r\left( \omega \right) _{11}^{-1}-\Sigma ^r\left( \omega
\right) _{11}][g^r\left( \omega \right) _{22}^{-1}-\Sigma ^r\left( \omega
\right) _{22}]-\Sigma ^r\left( \omega \right) _{12}\Sigma ^r\left( \omega
\right) _{21}\text{ }
\end{equation}

\bigskip

{\it Choice of Integration Contour.} The supercurrent formula Eq.[6] can be
expressed as 
\begin{equation}
I_{sc}=-\frac{4e\Gamma }{\hbar }%
\mathop{\rm Re}%
\{\int_{-\infty }^{+\infty }\frac{d\omega }{2\pi }[G_{d}^{r}\left( \omega
\right) _{12}g_{L}^{<}\left( \omega \right) _{21}+G_{d}^{<}\left( \omega
\right) _{12}g_{L}^{a}\left( \omega \right) _{21}]\}\text{ .}
\end{equation}
In order to avoid calculating this integration numerically, we shall
transform it into a contour integral. First, using the
fluctuation-dissipation theorem $G^{<}\left( \omega \right) =[G^{a}\left(
\omega \right) -G^{r}\left( \omega \right) ]f(\omega )$, where $G$ stands
for any Green function in the equilibrium state and $f(\omega )$ is the
Fermi distribution function, we have \ \ \ 
\begin{equation}
I_{sc}=\frac{4e\Gamma }{\hbar }%
\mathop{\rm Re}%
\{\int_{-\infty }^{+\infty }\frac{d\omega }{2\pi }[G_{d}^{r}\left( \omega
\right) _{12}g_{L}^{r}\left( \omega \right) _{21}f(\omega )-G_{d}^{a}\left(
\omega \right) _{12}g_{L}^{a}\left( \omega \right) _{21}f(\omega )]\}\text{ .%
}
\end{equation}
Then we make a change of the integration variable $\omega \rightarrow \omega
-i0^{+}$ for $G_{d}^{r}\left( \omega \right) _{12}g_{L}^{r}\left( \omega
\right) _{21}f(\omega )$ and $\omega \rightarrow \omega +i0^{+}$ for $%
G_{d}^{a}\left( \omega \right) _{12}g_{L}^{a}\left( \omega \right)
_{21}f(\omega )$ to divide $I_{sc}$ into two integrals along different
paths: 
\begin{multline}
I_{sc}=\frac{4e\Gamma }{\hbar }%
\mathop{\rm Re}%
\{\int_{-\infty +i0^{+}}^{+\infty +i0^{+}}\frac{d\omega }{2\pi }%
[G_{d}^{r}\left( \omega -i0^{+}\right) _{12}g_{L}^{r}\left( \omega
-i0^{+}\right) _{21}f(\omega -i0^{+})]  \nonumber \\
-\int_{-\infty -i0^{+}}^{+\infty -i0^{+}}[G_{d}^{a}\left( \omega
+i0^{+}\right) _{12}g_{L}^{a}\left( \omega +i0^{+}\right) _{21}f(\omega
+i0^{+})]\}\text{ .}
\end{multline}
Since the above two integrands are actually the same, we define a new
integrand equal to them: 
\begin{eqnarray}
J(\omega ) &\equiv &G_{d}^{r}\left( \omega -i0^{+}\right)
_{12}g_{L}^{r}\left( \omega -i0^{+}\right) _{21}f(\omega -i0^{+})  \nonumber
\\
&=&G_{d}^{a}\left( \omega +i0^{+}\right) _{12}g_{L}^{a}\left( \omega
+i0^{+}\right) _{21}f(\omega +i0^{+})
\end{eqnarray}
Thus the current integration can be rewritten into a contour integral in the
complex-$\omega $ plane: 
\begin{equation}
I_{sc}=\frac{4e\Gamma }{\hbar }%
\mathop{\rm Re}%
[(\int_{-\infty +i0^{+}}^{+\infty +i0^{+}}-\int_{-\infty -i0^{+}}^{+\infty
-i0^{+}})\frac{d\omega }{2\pi }J(\omega )]\text{ }=\frac{4e\Gamma }{\hbar }%
\mathop{\rm Re}%
\int_{c}J(\omega )\frac{d\omega }{2\pi }\text{.}
\end{equation}
The integration path $C$ = $\int_{-\infty +i0^{+}}^{+\infty
+i0^{+}}+\int_{+\infty -i0^{+}}^{-\infty -i0^{+}}$ is a close contour which
lies infinitely close to the real $\omega $ axis. The integral can then be
evaluated analytically using Cauchy's theorem.

This choice of the integration contour has the advantage of leaving all
(infinitely many) the poles at $\omega =i\frac{(2n+1)\pi }{\beta }$\ arising
from $f(\omega )$ outside the contour, thus avoid the calculation of the
summation over all the Matsubara frequencies $(\omega =i\frac{(2n+1)\pi }{%
\beta })$.

{\it Poles and definition of Andreev bound state. }Since the current is
proportional to the sum of the residues of the integrand $J(\omega )$ inside
the contour $C$, only real poles of $J(\omega )$ contribute. For $J(\omega
)\equiv G_{d}^{r}\left( \omega -i0^{+}\right) _{12}g_{L}^{r}\left( \omega
-i0^{+}\right) _{21}f(\omega ),$ there are only two pairs of real poles: $%
\omega =\pm \Delta $ and $\omega =\pm \epsilon ^{\ast }(\left| \epsilon
^{\ast }\right| <\Delta )$. The poles at $\omega =\pm \Delta $ originate
from the singularities of the DOS of the BCS superconductors.

The poles at $\omega =\pm \epsilon ^{\ast }$ come from the QD's Green
function and are two real roots of the equation $D(\omega )=0$ inside the
gap ( $\left| \epsilon ^{\ast }\right| <\Delta $ ). These two poles give the
positions of the quasi-particle states of the QD inside the gap. The fact
that they are real indicates that these two quasi-particle states are
exactly bound states, usually called the Andreev bound states (ABS).

Before coupling to the superconductors, the unperturbed QD has a bound state
at $\epsilon _0$ for the electron spectrum and one at $-\epsilon _0$\ for
the hole spectrum. After coupled to the superconductors, through the
electron-transfer with the two superconductors via Andreev process, the
electron part and the hole part of the QD's Green function are coupled
together, and give rise to significant modifications to the QD's energy
spectrum: the original bound state at $\epsilon _0$ for electron (or $%
-\epsilon _0$ for hole) renormalizes into two symmetrical\ bound states at $%
\pm \epsilon ^{*}$ for both electron and hole spectrum.

To express $\epsilon ^{*}$ as the algebraic function of $\epsilon _0$, $\phi 
$ and $\Gamma $, we transform the equation $D(\omega )=0$ into a quartic
equation of $x=$ $\epsilon ^{*2}$: 
\begin{gather}
x^4-(2A^2-4\Gamma ^2)x^3+(A^4+2\Delta ^2B^2-4\Gamma ^2\Delta
^2)x^2+(2A^2\Delta ^2B^2)x+(\Delta ^2B^2)^2=0;  \nonumber \\
A^2=\Delta ^2+\epsilon _0^2+\Gamma ^2,\text{\ }B^2=\epsilon _0^2+\Gamma
^2\cos ^2\frac \phi 2\text{ };\text{ \ \ \ }x=\epsilon ^{*2}<\Delta ^2.
\end{gather}
This quartic equation of $x$ has only one positive real solution $x_{0\text{ 
}}$less than $\Delta ^2$: $\pm \epsilon ^{*}=\pm \sqrt{x_0}.$ The quartic
equation is easy to solve algebraically; but we omit the lengthy solution.
The dependence of $\epsilon ^{*}$ on $\epsilon _0$, $\phi $ and $\Gamma $ is
shown graphically in \cite{Zhu2001}.

\bigskip

{\it Residues and analytical result of the current. }The current is obtained
by adding up the residues of the four poles: 
\begin{equation}
I_{sc}=\frac{4e\Gamma }\hbar 
\mathop{\rm Im}%
[\sum_{i=1}^4%
\mathop{\rm Re}%
sJ(\omega _i)],\text{ }
\end{equation}
where $\omega _i=\pm \epsilon ^{*},\pm \Delta .$ The residues of $J$ at $%
\epsilon ^{*}$ and $-\epsilon ^{*}$ are : 
\begin{gather}
\mathop{\rm Re}%
\text{s}_{\omega =\pm \epsilon ^{*}}J(\omega )=\lim_{\omega \rightarrow \pm
\epsilon ^{*}}[\frac{\omega -(\pm \epsilon ^{*})}{D(\omega )}\Sigma ^r\left(
\omega -i0^{+}\right) _{12}g_L^r\left( \omega -i0^{+}\right) _{21}f(\omega
)], \\
\mathop{\rm Im}%
[%
\mathop{\rm Re}%
\text{s}_{\omega =\pm \epsilon ^{*}}J(\omega )]=\mp \frac \Gamma 2\frac{%
\Delta ^2}{\Delta ^2-\epsilon ^{*2}}\frac{f(\pm \epsilon ^{*})\sin \phi }{%
2\epsilon ^{*}[(\Delta ^2-\epsilon ^{*2})+\frac \Gamma {\sqrt{\Delta
^2-\epsilon ^{*2}}}(2\Delta ^2-\epsilon ^{*2})+\frac{\Gamma ^2}{\Delta
^2-\epsilon ^{*2}}\Delta ^2\sin ^2\frac \phi 2]}.  \nonumber
\end{gather}
\newline
The residues at $\pm \Delta $ turn out to be zero since 
\begin{equation}
D(\pm \Delta )=\infty ,\ \ \ \ \ \ \ \frac 1{D(\pm \Delta )}=0.
\end{equation}
Thus the supercurrent is simply 
\begin{equation}
I_{sc}=\frac{e\Gamma ^2}\hbar \frac{\Delta ^2}{\Delta ^2-\epsilon ^{*2}}%
\frac{\sin \phi \tanh \frac{\beta \epsilon ^{*}}2}{\epsilon ^{*}[(\Delta
^2-\epsilon ^{*2})+\frac \Gamma {\sqrt{\Delta ^2-\epsilon ^{*2}}}(2\Delta
^2-\epsilon ^{*2})+\frac{\Gamma ^2}{\Delta ^2-\epsilon ^{*2}}\Delta ^2\sin ^2%
\frac \phi 2]}\text{.}
\end{equation}
where $\phi =\phi _R-\phi _L$ is the phase difference of the two
superconductors.

{\it Discussion. }In addition to the splitting of the original energy level $%
\epsilon _0$ into the two ABS, another modification to the QD's energy
spectrum is that the QD begins to have a continuum energy spectrum outside
the gap. In fact, one way to calculate the current integral ( $I_{sc}=-\frac{%
4e\Gamma }\hbar 
\mathop{\rm Re}%
\{\int_{-\infty }^{+\infty }\frac{d\omega }{2\pi }[G_d^r\left( \omega
\right) _{12}g_L^{<}\left( \omega \right) _{21}+G_d^{<}\left( \omega \right)
_{12}g_L^a\left( \omega \right) _{21}]\}$ ) is to divide the QD's spectrum
into the continuum part ( $\left| \omega \right| >\Delta $ ) and the
discrete part ( $\left| \omega \right| <\Delta $ ), and then evaluate their
contributions to the current separately. However, this method involves a
numerical integration and faces the problem of explaining why the peaks at $%
\omega =\pm \Delta $ in $I_{continuum}$ vs. $\epsilon _0$ curve and in $%
I_{discrete}$ vs. $\epsilon _0$ curve exactly cancel out when the two parts
add up to give the total current, see Fig.1.

In contrast, from our approach, this exact cancellation can be shown
naturally. The current is obtained by adding the residues of the four poles.
Since the residues at $\pm \Delta $ vanish, the total current curve has no
peaks arising from the DOS singularities of the superconductors.

Now only residues at $\pm \epsilon ^{*}$ contribute to the total current and
give rise to a sharp peak in $I$ vs. $\epsilon _0$ curve at $\epsilon _0=0.$
The reason for the appearance of the peak at $\epsilon _0=0$ is that when $%
\epsilon _0=0$, the upper and the lower ABS have same phase. As $\epsilon _0$
departs from the Fermi level, the phase of the upper and lower ABS begins to
differ and this difference severely reduces the supercurrent across the
junction. Fig.2 shows the central peaks at different $\Gamma ,$ the peaks
broaden when $\Gamma $ increases.

A direct application of this result is a way to align the energy level of
the QD to the Fermi level of the superconductor in the actual experiment:
when changing the gate voltage, the state with the maximum supercurrent is
at $\epsilon _{0}=0.$

{\it Conclusion. }In summary, we have calculated analytically the
supercurrent across an S-QD-S junction. The central results are Eq.[18] and
Eq.[21]. They give the analytical result of both the positions of ABS and
the supercurrent. We show two additional advantages of our method: (1) It
presents a natural definition of the Andreev bound states inside the QD. (2)
It explains the disappearance of the peaks at the gap edge in $I$ vs. $%
\epsilon _0$ curve.\bigskip \ The method devised in the paper can also be
used in computing other equilibrium properties of the superconductor-normal
hybrid systems.

We thank Q.F. Sun for his stimulating remarks. This work is supported by
National Science Foundation of China under the Grant No. 10074001.


\newpage

\section*{Figure Captions}

\begin{itemize}
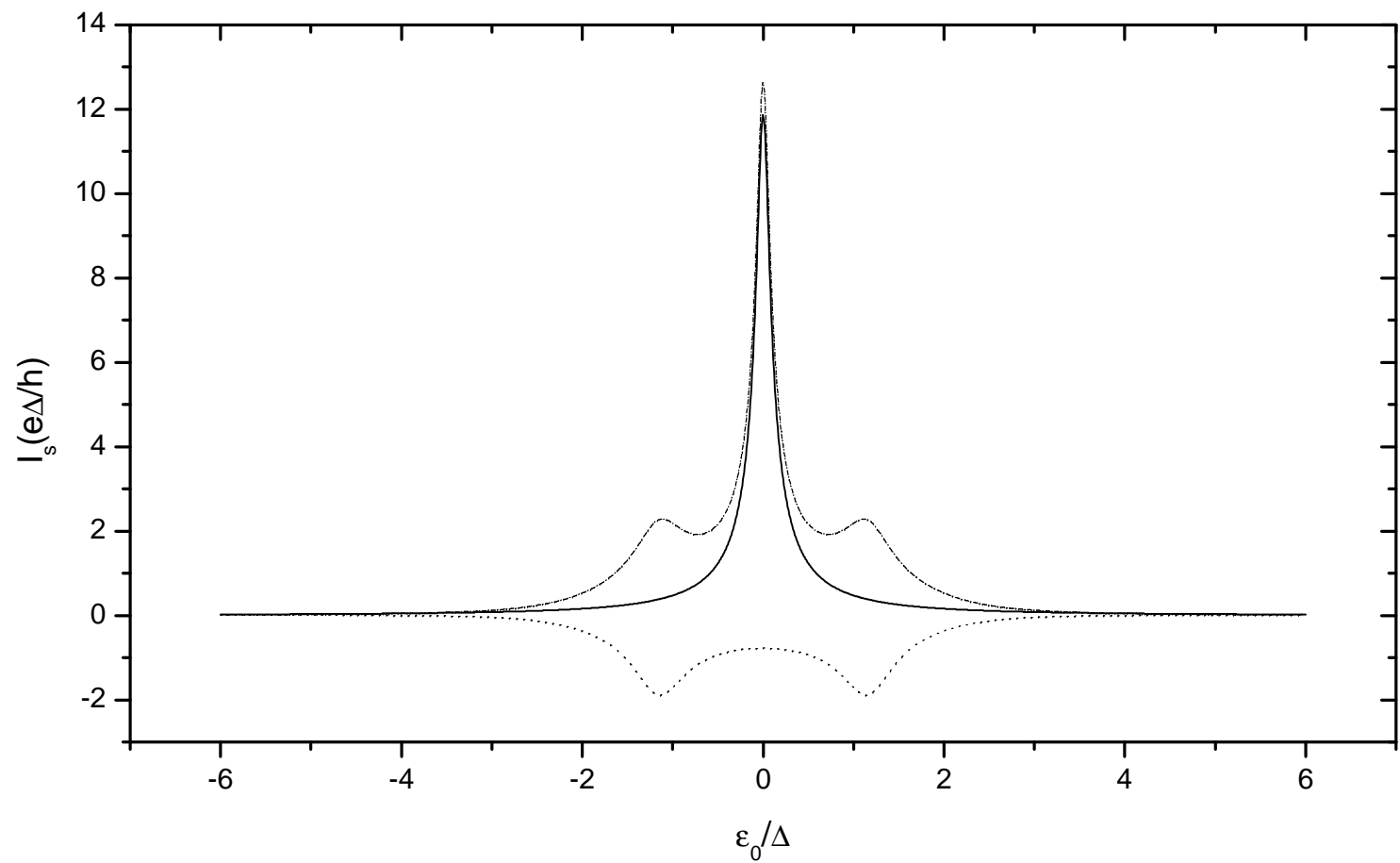
\item[{\bf Fig. 1}]  Supercurrent vs. $\epsilon _0$ when $\Gamma =0.1$ and $%
\phi =\frac \pi 2.$ The dash line is from discrete spectrum, and the dot
line from continuum spectrum. The solid line is total current. Note that the
continuum spectrum contributes negative current.

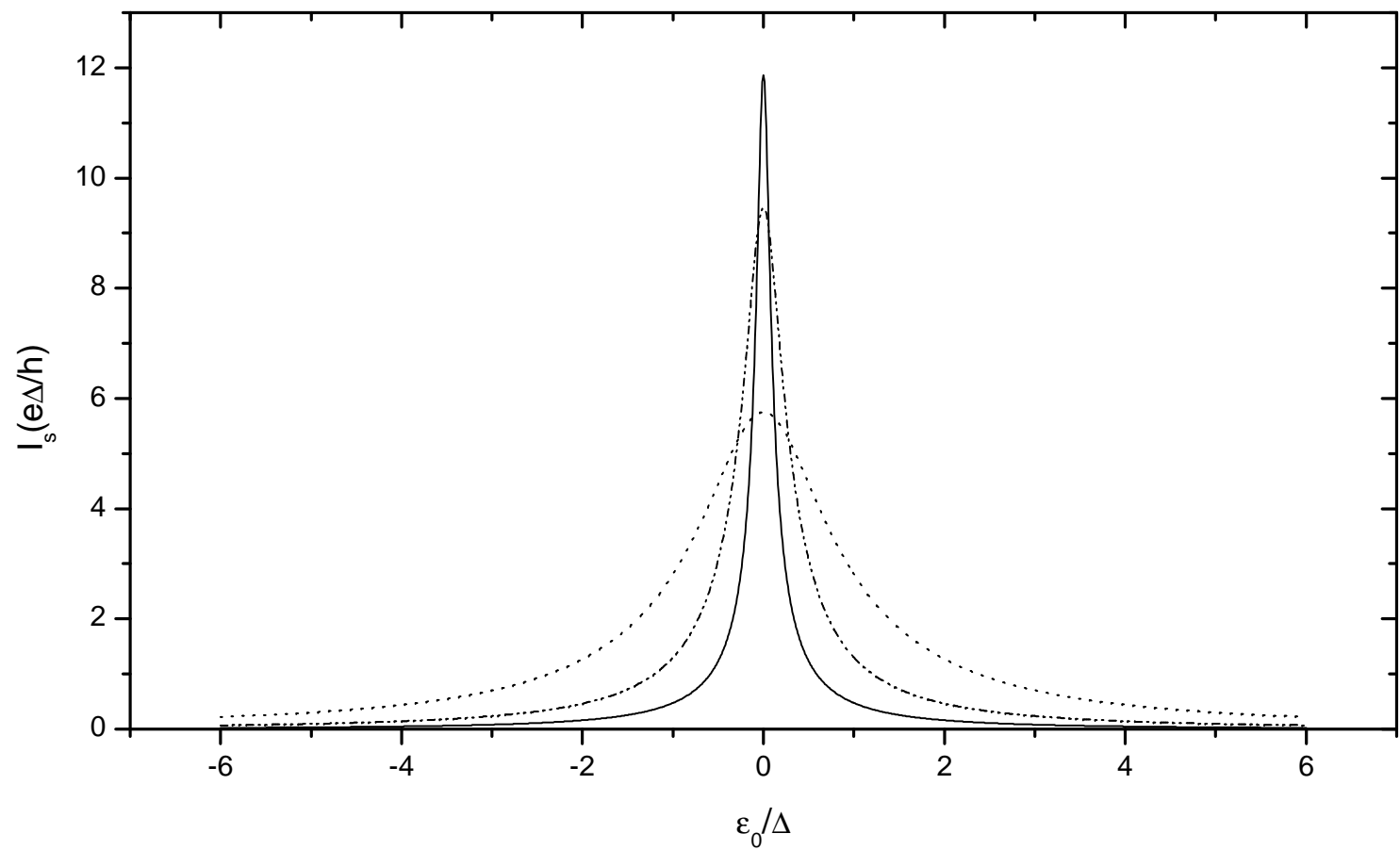
\item[{\bf Fig. 2}]  Supercurrent vs. $\epsilon _0$ with $\phi =\frac \pi 2$
for different $\Gamma .$ Different curves are for $\Gamma =0.1,$ $0.3,$ $1.0,
$ respectively. Note the broadening effect as $\Gamma $ increases.
\end{itemize}

\end{document}